\documentclass[11pt,reqno,final]{article}
\usepackage{amsthm}
\usepackage{amsmath,amsfonts,amssymb,amsthm,version}
\usepackage{mathrsfs,fancybox,pifont}
\usepackage{graphicx}
\usepackage{url,hyperref}
\usepackage[notcite,notref]{showkeys}
\usepackage{color}
\usepackage{multirow}
\usepackage{epstopdf}
\usepackage{cases}
\usepackage{mathtools}
\usepackage{algorithm,algorithmic}
\usepackage{authblk}
\usepackage{fancyhdr}
\usepackage{lipsum}
\usepackage{cite}
\usepackage{float}
\usepackage[top=2cm]{geometry}
\usepackage{booktabs}
\usepackage{subcaption}  
\usepackage[table]{xcolor}

\allowdisplaybreaks

\numberwithin{equation}{section}
\numberwithin{figure}{section}
\numberwithin{table}{section}

\theoremstyle{plain}

\theoremstyle{definition}

\theoremstyle{remark}

\title{BioSEN: A Bio-acoustic Signal Enhancement Network for Animal Vocalizations}
\author[1]{Tianyu Song} 
\author[1,2]{Ton Viet Ta} 
\author[3]{Ngamta Thamwattana} 
\author[2]{Hisako Nomura}
\author[4]{Linh Thi Hoai Nguyen}

\affil[1]{Graduate School of Bioresource and Bioenvironmental Science, Kyushu University}
\affil[2]{Faculty of Agriculture, Kyushu University}
\affil[3]{School of Information and Physical Sciences, University of Newcastle}
\affil[4]{International Institute for Carbon-Neutral Energy Research, Kyushu University}

\date{}

\begin{document}
%
\maketitle
\begin{abstract}
Most work in audio enhancement targets human speech, while bioacoustics is less studied due to noisy recordings and the distinct traits of animal sounds. To fill this gap, we adapt speech enhancement methods and build BioSEN, a model made for bioacoustic signals. BioSEN has three modules: a multi-scale dual-axis attention unit for time–frequency feature extraction, a bio-harmonic multi-scale enhancement unit for capturing harmonic structures, and an 
 energy-adaptive gating connection unit that uses frequency weights to keep vocalizations from being removed as noise. Tests on three bioacoustic datasets show that BioSEN matches or exceeds state-of-the-art speech enhancement models while using far less computation. These results show BioSEN’s strength for bioacoustic audio enhancement and its promise for biodiversity monitoring and conservation.
 
 
\end{abstract}

\section{Introduction}\label{sec:intro}
With the rapid growth of environmental challenges and biodiversity loss, ecological monitoring has become increasingly important for conservation and ecosystem management. Among existing monitoring approaches, bioacoustics has emerged as a powerful and non-invasive tool for studying wildlife populations and biodiversity.  {In particular, passive acoustic monitoring and computational bioacoustics have become important research directions because they allow long-term, large-scale, and low-disturbance monitoring of animals in natural habitats \cite{sugai2019terrestrial,stowell2022computational}.} Acoustic signals can be used to identify species \cite{kohlberg2024buzzes}, estimate population density \cite{navine2024counting}, and support biodiversity conservation in environments where visual observation is difficult or impractical \cite{rasmussen2024sound}.  {Recent deep learning systems such as BirdNET have further demonstrated the potential of AI-based acoustic analysis for biodiversity monitoring \cite{kahl2021birdnet}.}  

Despite its advantages, the effectiveness of bioacoustic monitoring strongly depends on the quality of recorded audio signals. Because most recordings are collected in natural environments, bioacoustic data are often contaminated by environmental noise such as wind, rain, flowing water, insect sounds, and overlapping vocalizations from multiple species. These factors typically result in recordings with very low signal-to-noise ratios (SNRs), making it difficult for artificial intelligence (AI) systems to accurately detect and analyze target animal vocalizations \cite{sharma2023review}. Consequently, robust signal enhancement and denoising methods are essential for reliable bioacoustic analysis.  {Therefore, bioacoustic enhancement can be regarded as an important intermediate task between field audio collection and downstream ecological inference.}  

Existing research on acoustic enhancement has primarily focused on human speech \cite{gajecki2025adversarial,dementyev2025submillisecond} and underwater acoustic signals \cite{zhao2025dual,tang2025novel}. These studies have introduced effective architectures and learning strategies for suppressing noise while preserving target signals.  {They provide useful technical foundations, including complex-valued spectral modeling, attention mechanisms, and multi-scale feature extraction.} However, animal vocalizations differ substantially from human speech in terms of frequency distribution, harmonic organization, temporal sparsity, and acoustic variability \cite{barnhill2024animal}. For example, many bird and animal calls exhibit narrow-band harmonic structures and intermittent temporal patterns that are not commonly observed in speech signals. Furthermore, bioacoustic recordings often contain rapidly changing broadband environmental noise \cite{juodakis2022wind,song2025}, which further complicates the enhancement task. As a result, speech-oriented enhancement models may not effectively generalize to bioacoustic applications.  {This gap motivates the development of enhancement models that are not only adapted from speech processing, but also explicitly designed around bioacoustic signal characteristics.}  

Another major challenge in bioacoustic enhancement is the scarcity of clean reference recordings. Unlike speech enhancement research, where large-scale clean datasets are readily available, obtaining truly noise-free animal vocalizations in natural habitats is extremely difficult. To address this limitation, recent studies have proposed pseudo-clean training strategies for bioacoustic denoising \cite{miron2024biodenoising,sarkar2025comparing}. These approaches demonstrate that speech pre-trained models can be adapted to generate pseudo-clean references for training bioacoustic enhancement systems.  {Thus, recent bioacoustic denoising research has begun to move from direct application of speech enhancement models toward bioacoustic-specific training and model design.} Inspired by these advances, we also adopt a pseudo-clean data generation strategy based on a human-speech pre-trained enhancement model.  

Motivated by the unique characteristics of animal vocalizations and the limitations of existing approaches, we propose BioSEN (Bio-acoustic Signal Enhancement Network), a lightweight neural network specifically designed for bioacoustic signal enhancement.  {The proposed method is positioned at the intersection of computational bioacoustics and deep learning-based audio enhancement: it uses ideas from speech and underwater acoustic enhancement, while introducing modules tailored to harmonic and frequency-dependent animal vocalizations.} BioSEN integrates specialized modules to model harmonic structures, multi-scale spectral patterns, and adaptive feature transmission for animal vocalizations. In particular, the proposed framework is designed to preserve biologically meaningful acoustic structures while effectively suppressing environmental noise under computationally efficient settings.  

Motivated by the unique characteristics of animal vocalizations and the limitations of existing approaches, we propose BioSEN (Bio-acoustic Signal Enhancement Network), a lightweight neural network specifically designed for bioacoustic signal enhancement. BioSEN integrates specialized modules to model harmonic structures, multi-scale spectral patterns, and adaptive feature transmission for animal vocalizations. In particular, the proposed framework is designed to preserve biologically meaningful acoustic structures while effectively suppressing environmental noise under computationally efficient settings.  

The main contributions of this work are summarized as follows:

\begin{enumerate}
    \item We adopt the complex spatial coordinate convolution autoencoder (CSCConv-AE) \cite{tang2025novel} as the baseline framework and demonstrate through ablation experiments that it provides substantial improvements in denoising performance, establishing an effective backbone for bioacoustic enhancement.  
    
    \item We propose a multi-scale dual-axis attention (MSDA) module that jointly models temporal, spectral, and channel-wise dependencies, enabling the network to better focus on target animal vocalizations while suppressing irrelevant noise components.  
    
    \item We introduce a bio-harmonic multi-scale enhancement (BHME) module that employs frequency-axis multi-scale convolutions to capture harmonic structures associated with different fundamental frequencies, thereby improving the preservation of biologically relevant acoustic patterns.  
    
    \item We develop an energy-adaptive gating connection (EAGC) module that combines frequency-aware gating and cross-attention mechanisms to selectively transfer informative encoder features to the decoder, preserving spectral and phase information while reducing noise propagation.  
    
    \item Experimental results on multiple bioacoustic datasets demonstrate that BioSEN consistently achieves competitive or superior enhancement performance compared with state-of-the-art speech enhancement models, while requiring significantly lower computational complexity.
\end{enumerate}

The remainder of this paper is organized as follows. Section~\ref{sec:method} describes the proposed BioSEN framework, including the MSDA, BHME, and EAGC modules, as well as the datasets and experimental settings. Section~\ref{sec:result} presents the experimental results, ablation studies, and comparisons with existing speech enhancement models. Finally, Section~\ref{sec:con} concludes the paper and discusses future research directions.

\section{Materials and Methods}
\label{sec:method}
\subsection{Overall Architecture}
\begin{figure}[htb]
    \begin{minipage}[b]{1.0\linewidth}
        \centering
        \centerline{\includegraphics[scale=0.7]{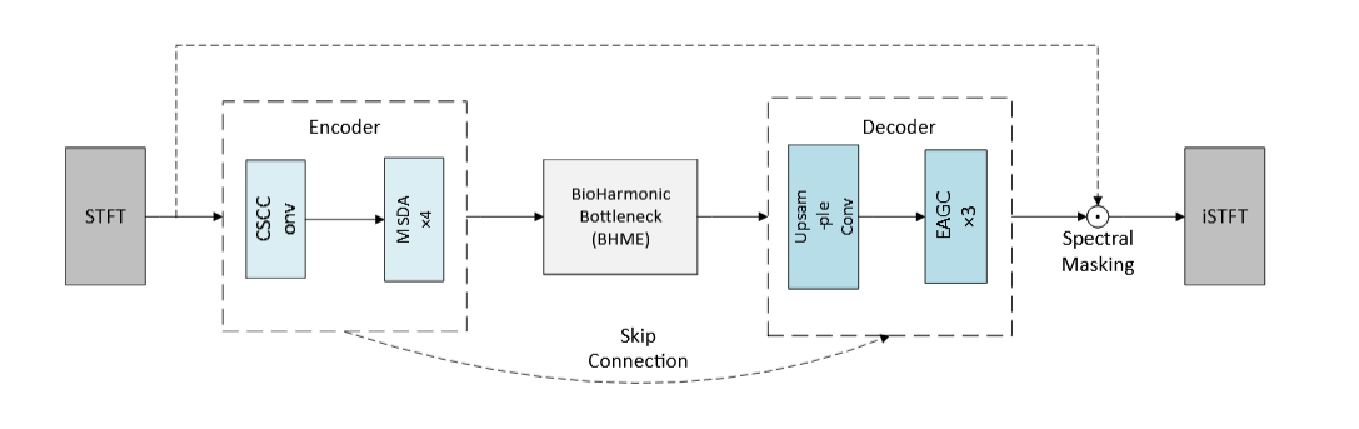}}
    \end{minipage}
    \caption{Architecture of BioSEN.}
    \label{fig:model} 
\end{figure}

Animal vocalizations differ fundamentally from human speech and broadband noise. For example, bird calls often concentrate energy in narrow frequency bands, forming harmonic structures aligned with fundamental frequencies, while chirps or calls are separated by silent intervals, creating sparse temporal patterns unlike continuous speech. To capture these unique properties, we design BioSEN, whose overall structure is shown in Figure~\ref{fig:model}. The framework integrates three core modules: multi-scale dual-axis attention (MSDA), bio-harmonic multi-scale enhancement (BHME), and energy-adaptive gating connection (EAGC), which are described in the following subsections.  

\subsection{Multi-Scale Dual-Axis Attention (MSDA)}
\begin{figure}[htb]
    \begin{minipage}[b]{1.0\linewidth}
        \centering
        \centerline{\includegraphics[scale=0.7]{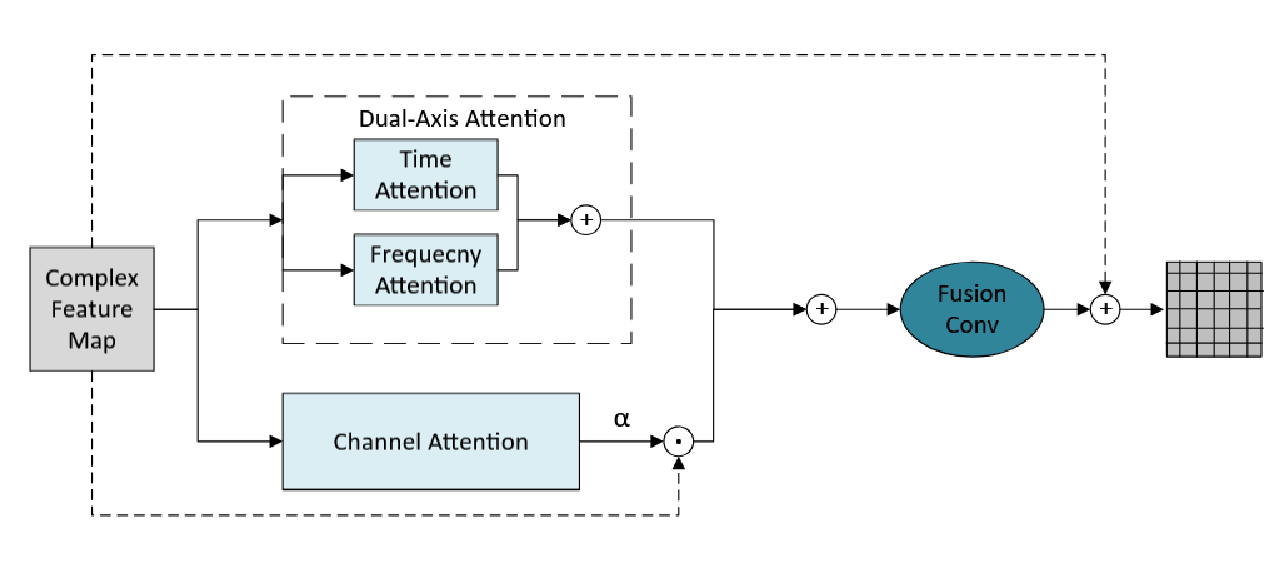}}
    \end{minipage}
    \caption{Structure of  the MSDA module.}
    \label{fig:msda} 
\end{figure}

To model the complex time–frequency structure of bioacoustic signals, we introduce the MSDA module as the core component of the encoder. MSDA uses a bifurcated design that separately captures temporal and spectral dependencies while also incorporating inter-channel attention, enabling the network to highlight biologically meaningful patterns (Figure~\ref{fig:msda}).  

Given complex input features $X \in \mathbb{R}^{B \times F \times T \times C \times 2}$, the process is as follows. First, dual multi-head attention layers capture temporal and frequency contextual information ($A_{\text{dual}}$): Specifically, the input X is reshaped to (B×F,T,C) for time attention and (B×T,F,C) for frequency attention to capture respective dependencies. This axis-specific separation is designed to improve the ability to analyze animal vocalizations.  

Next, channel attention evaluates the relative importance of each feature channel. Specifically, the magnitude $M$ of $X$ is globally pooled and passed through a fully connected layer with sigmoid activation $\sigma$ to produce a channel weight vector $\alpha$, which adaptively emphasizes features correlated with bioacoustic patterns.

The output combines dual-axis and channel attention:
\begin{equation}
X_{\text{out}} = X + \text{Conv}_{1\times1}\big( A_{\text{dual}}(X) + (X \odot \alpha) \big),
\end{equation}
where the $1 \times 1$ complex convolution fuses the two branches, and the residual connection stabilizes training while preserving input information. Here, the operator $\odot$ denotes the Hadamard product, i.e., element-wise multiplication. This operation scales each element of $X$ by its corresponding channel weight in $\alpha$ (with broadcasting applied as needed).

\subsection{Bio-Harmonic Multi-Scale Enhancement (BHME)}
\begin{figure}[htb]
    \begin{minipage}[b]{1.0\linewidth}
        \centering
        \centerline{\includegraphics[scale=0.7]{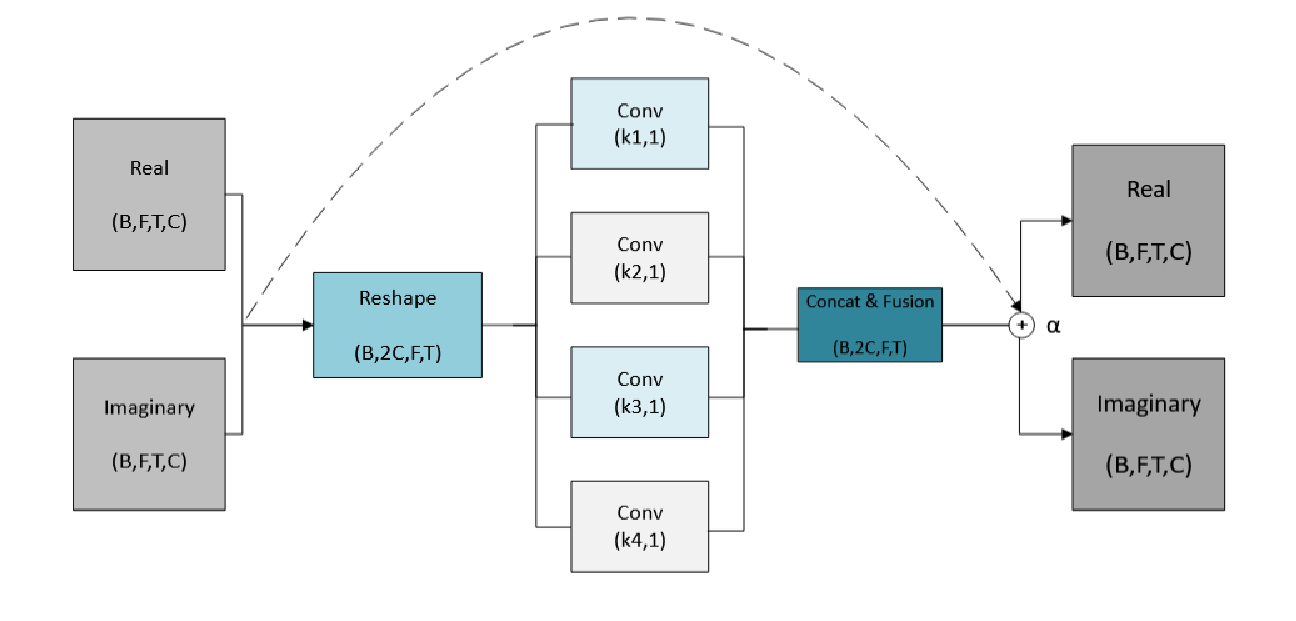}}
    \end{minipage}
    \caption{Structure of  the BHME module.}
    \label{fig:bh} 
\end{figure}

In conventional speech research, little distinction is made between the fundamental frequency and its higher harmonics. In contrast, bioacoustic signals rely heavily on these structures. For instance, a bird call may exhibit a fundamental frequency rising from 1 kHz to 1.5 kHz, with its second harmonic shifting from 2 kHz to 3 kHz, and so on. Such harmonic patterns directly shape the perceived “crispness” or “sharpness” of the call. To capture this property, we design a bottleneck-layer module using learnable anisotropic kernels (k×1) to enhance harmonic features, as illustrated in Figure~\ref{fig:bh}.

Unlike fixed Gammatone filters, this module naturally learns diverse harmonic spacing patterns via gradient optimization without explicit constraints. Using a parallel multi-branch structure, it mimics the auditory perception of animal calls with varying fundamental frequencies and pitches. This allows the model to emphasize harmonic patterns of different densities, preserving key signal characteristics while reducing confusion between harmonic details and noise.  

Specifically, BHME captures multi-scale features by employing parallel anisotropic convolutional filters with different kernel sizes along the frequency axis. Each filter specializes in a particular harmonic spacing:  
\begin{equation}
H_k = \text{Conv}_{(k,1)}(X_{conv}).
\end{equation}
The resulting multi-scale features are concatenated and fused by a $1 \times 1$ convolution to form a unified harmonic representation $H_{fused}$, which is then combined with the original input via a scaled residual connection:  
\begin{equation}
Y = X + \alpha \cdot H_{fused}.
\end{equation}

Since harmonic structures are primarily distributed along the frequency axis, all four convolution kernels use a width of 1, ensuring that the model focuses on spectral features without unnecessary computation along the time axis. This design aims to enable parallel, asymmetric frequency analysis across multiple scales, thereby effectively capturing harmonic structures ranging from dense to sparse.

\subsection{Energy-Adaptive Gating Connection (EAGC)}
\begin{figure}[htb]
    \begin{minipage}[b]{1.0\linewidth}
        \centering
        \centerline{\includegraphics[scale=0.6]{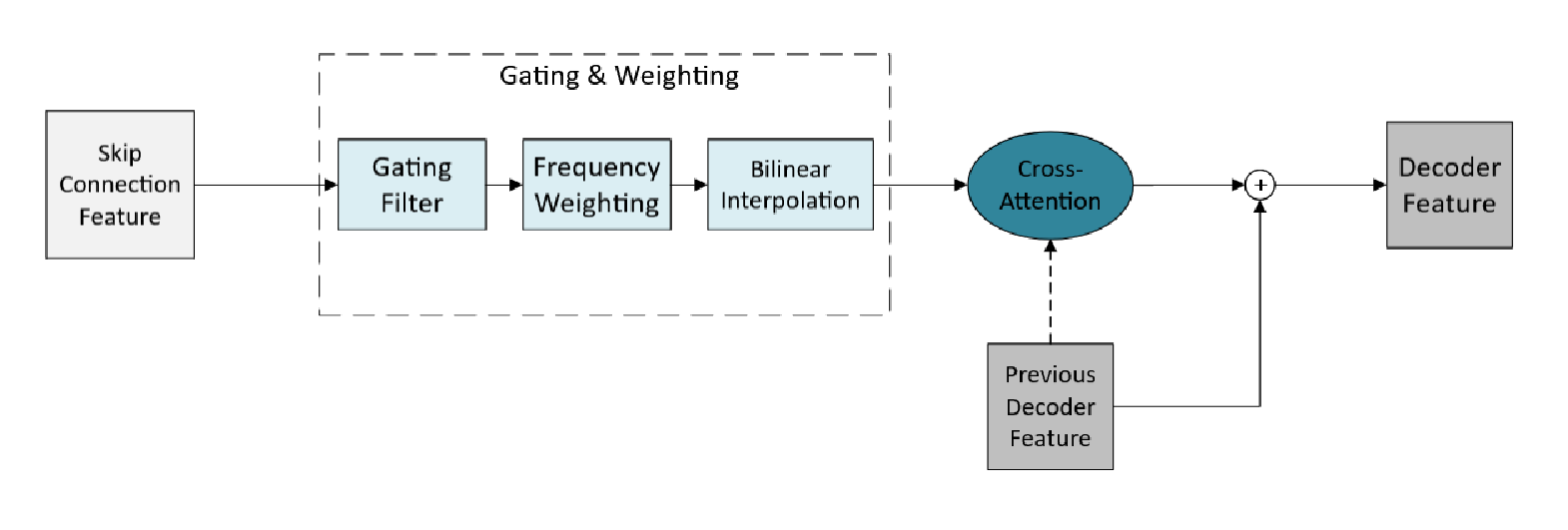}}
    \end{minipage}
    \caption{Structure of the EAGC module.}
    \label{fig:eagc} 
\end{figure}

Skip connections are essential for reconstruction but may also transfer noise. To address this, we propose the EAGC module, which applies a two-stage filtering process to selectively pass only the most informative features from encoder to decoder. The structure is shown in Figure~\ref{fig:eagc}.  

EAGC acts as an adaptive filter by combining frequency-aware gating with cross-attention. For encoder features $E$ and decoder features $D$, the process is as follows:  

\textit{Frequency-weighted gating:} An initial learnable gate $G$ identifies candidate encoder features $E_o$. These are then modulated by a frequency energy weight $W_{freq}$, derived from the spectral energy distribution to preserve harmonic bands. The gated features are:
\begin{equation}
E_f = E_o \odot \big(\sigma(\text{Conv}(E_o)) \cdot W_{freq}\big).
\end{equation}

\textit{Cross-attention:} The decoder state $D$ serves as a query to select the most relevant features from $E_f$, producing the refined representation $E_s$. These features are concatenated with $D$ to enhance reconstruction while suppressing noise.  

To handle resolution mismatches between encoder and decoder layers, EAGC applies bilinear interpolation:
\begin{equation}
E_a = \text{Interpolate}(E_f, \text{size} = (F_d, T_d)),
\end{equation}
which ensures that features from different encoder layers can be effectively integrated with their corresponding decoder layers, regardless of spatial resolution differences.

Through this design, EAGC effectively transmits biologically relevant features while minimizing noise propagation in skip connections.

\subsection{Dataset}
As noted in Section~\ref{sec:intro}, most bioacoustic recordings are made in the field and include substantial environmental noise. To reduce this, we use the pseudo-clean target method from \cite{miron2024biodenoising}, which turns noisy recordings into paired samples of the original input and a pseudo-clean reference. Using this, we convert the Xeno Bird dataset \cite{vellinga2025xenocanto} into a training set of noisy/pseudo-clean pairs.

Because truly clean bioacoustic recordings remain scarce, we further evaluate our model on three small but diverse test sets to reduce the bias introduced by limited samples. A summary of all datasets is provided in Table~\ref{tab:data}.

\begin{table}[!th]
\renewcommand{\arraystretch}{1.0}
\caption{Dataset description.}
\label{tab:data}
\centering
\resizebox{\columnwidth}{!}{%
\begin{tabular}{|l|cccc|}
\hline
Data Source & Xeno Bird\cite{vellinga2025xenocanto} & Bird Song\cite{earthspecies2020library} & Biodenoising\cite{miron2024biodenoising} & Mixed data\cite{Mumm2014,Elie2018,Yin2004}  \\
\hline
Usage & Train & Test & Test & Test \\
Animal species & Birds & Birds &  Chicken, Lion, etc. & Fruit Bats, Otters, etc. \\
SNR(Avg.) & - & [-10,-5] &  [-10,-5] & [-5,10] \\
\hline
\end{tabular}%
}
\end{table}



\subsection{Evaluation Metrics}
We evaluate model performance using perceptual quality, signal fidelity, and computational efficiency metrics.  

Scale-Invariant Signal-to-Distortion Ratio (SI-SDR): 
This measures the similarity between the enhanced signal $\hat{s}$ and the reference clean signal $s$, while being invariant to scale:
\begin{equation*}
s_{\text{target}} = \frac{\langle \hat{s}, s \rangle}{\| s \|^2} s,
\quad
e_{\text{noise}} = \hat{s} - s_{\text{target}},
\end{equation*}
\begin{equation*}
\text{SI-SDR} = 10 \log_{10} \frac{\| s_{\text{target}} \|^2}{\| e_{\text{noise}} \|^2}.
\end{equation*}
Higher SI-SDR indicates better preservation of the target signal.  

SI-SDR Improvement (SI-SDRi):
SI-SDRi quantifies the gain over the noisy input $x$, defined as:
\begin{equation*}
\text{SI-SDRi} = \text{SI-SDR}(\hat{s}, s) - \text{SI-SDR}(x, s).
\end{equation*}

Signal-to-Noise Ratio (SNR): 
SNR evaluates the ratio between clean signal power and residual noise power:
\begin{equation*}
\text{SNR} = 10 \log_{10} \frac{\| s \|^2}{\| s - \hat{s} \|^2}.
\end{equation*}

SNR Improvement (SNRi): 
Similar to SI-SDRi, SNRi measures improvement relative to the noisy input:
\begin{equation*}
\text{SNRi} = \text{SNR}(\hat{s}, s) - \text{SNR}(x, s).
\end{equation*}

Floating-Point Operations (FLOPs): 
FLOPs represent the number of arithmetic operations required during inference, reflecting computational complexity. Lower FLOPs correspond to more efficient models, which is critical for real-time or resource-constrained applications.  

Overall, higher SI-SDR, SI-SDRi, SNR, and SNRi values indicate better denoising performance, while lower FLOPs indicate greater efficiency.

\subsection{Environment and Parameters}
All experiments are implemented in PyTorch and run on an NVIDIA A100 GPU with 40 GB RAM. The training setup uses an initial learning rate of $1\times10^{-3}$ with a decay coefficient of 0.7, batch size of 16, and the negative SI-SDR as the loss function.  

\section{Results}
\label{sec:result}
We evaluate the proposed BioSEN model through ablation studies and comparisons with representative speech enhancement methods across three datasets.  

Table~\ref{tab2} presents results on the Bird Song dataset. 
The ablation study shows that the baseline CSCConv-AE provides only modest gains, while the addition of MSDA, BHME, and EAGC each brings noticeable improvements.  
Among all variants, BioSEN achieves the highest SNR (5.73 dB) and SNRi (13.54 dB), surpassing both its ablated versions and all competing speech enhancement models.  
For SI-SDR and SI-SDRi, BioSEN attains strong performance (3.47 dB and 11.27 dB, respectively), ranking second overall—slightly below CSCConv-AE+MSDA, which overfits SI-SDR at the cost of poor SNR.  
Importantly, BioSEN achieves this balance while requiring only 3.15 GFLOPs, making it substantially more efficient than high-performance models such as Demucs (23.78 GFLOPs), DCCRN (27.69 GFLOPs), and FullSubNet (93.82 GFLOPs).  
Its FLOPs rank second-lowest overall, larger only than the ultra-lightweight LiSenNet, but with far superior denoising performance.  
These results highlight BioSEN as both effective and computationally efficient.

\begin{table}[!t]
\renewcommand{\arraystretch}{1.1}
\caption{Ablation results and comparison with speech enhancement models on Bird Song \cite{earthspecies2020library}.}
\label{tab2}
\centering
\resizebox{\columnwidth}{!}{%
\begin{tabular}{|l|ccccc|}
\hline
Model & SI-SDR & SI-SDRi & SNR & SNRi & Flops(G) \\
\hline
Noisy & -7.80 & - & -7.81 & - & -\\
FSPEN\cite{yang2024fspen} & -0.64 & 7.16 & 2.43 & 10.24 & 6.61\\
LiSenNet\cite{yan2025lisennet} & -5.14 & 2.86 & 0.48 & 8.49 & \textbf{0.11}\\
Demucs\cite{defossez2020real} & 3.16 & 10.96 & 5.31 & 13.12 & 23.78\\
DCCRN\cite{hu2020dccrn} & 3.15 & 10.95 & 5.29 & 13.10 & 27.69\\
FullSubNet\cite{chen2022fullsubnet} & 2.76 & 10.56 & 5.20 & 13.02 & 93.82\\
CSCConv-AE & -0.06 & 7.74 & 3.23 & 11.04 & -\\
\qquad +MSDA & \textbf{4.02} & \textbf{11.82} & -4.26 & 3.56 & -\\
\qquad+BHME & 2.38 & 10.18 & 4.81 & 12.63 & -\\
\qquad+EAGC & 2.89 & 10.69 & 5.20 & 13.01 & -\\
BioSEN & \textbf{3.47} & \textbf{11.27} & \textbf{5.73} & \textbf{13.54} & \textbf{3.15} \\
\hline
\end{tabular}%
}
\end{table}

To test BioSEN beyond bird sounds, we evaluate it on two extra datasets: Biodenoising and Mixed data. Table~\ref{tab3} shows the results. BioSEN achieves the best scores across all metrics. The baseline speech models also work well: BioSEN and FullSubNet are close on Biodenoising, while BioSEN and DCCRN are similar on Mixed data. These results show two points: first, using speech-based models to make pseudo-clean training data is effective; second, BioSEN consistently beats these strong baselines, proving its strength for animal vocal enhancement and robustness across species.

\begin{table}[H]
\renewcommand{\arraystretch}{1.1}
\caption{Performance comparison on two additional datasets.}
\label{tab3}
\centering
\resizebox{\columnwidth}{!}{%
    \begin{tabular}{|l|cccc|cccc|}
    \hline
    \multirow{2}{*}{Model} & \multicolumn{4}{c|}{Biodenoising\cite{miron2024biodenoising}} & \multicolumn{4}{c|}{Mixed data\cite{Mumm2014,Elie2018,Yin2004}} \\
    \cline{2-9}
    & SI-SDR & SI-SDRi & SNR & SNRi & SI-SDR & SI-SDRi & SNR & SNRi \\
    \hline
    Noisy & -7.49 & - & -3.71 & - & 2.89 & - & 3.20 & - \\
    FSPEN & 6.89 & 14.38 & 5.20 & 8.91 & 12.15 & 9.26 & 10.27 & 7.07  \\
    LiSenNet & -1.42 & 8.49 & 2.53 & 8.63 & 2.87 & -0.02 & 4.62 & 1.42  \\
    Demucs & 7.97 & 15.47 & 6.38 & 10.10 & 12.83 & 9.94 & 12.55 & 9.35 \\
    DCCRN & 7.14 & 14.63 & 6.19 & 9.90 & 15.97 & 13.08 & 13.09 & 9.40  \\
    FullSubNet & 9.17 & 16.66 & 6.43 & 10.14 & 13.89 & 11.00 & 13.82 & 10.62  \\
    BioSEN & \textbf{9.44} & \textbf{16.93} & \textbf{6.52} & \textbf{10.23} & \textbf{16.16} & \textbf{13.27} & \textbf{16.10} & \textbf{12.90}  \\
    \hline
    \end{tabular}
} 
\end{table}

\section{Conclusion}
\label{sec:con}

In this study, we present BioSEN, a denoising model built for biological audio signals. Based on traits of animal calls,  we design a harmonic monitoring and enhancement module, along with a frequency energy protection module that adapts to diverse vocalization patterns. These components enable the model to detect and preserve both continuous and transient signals, maintaining the integrity of the target audio. Experimental results on three datasets show that BioSEN matches or surpasses advanced speech enhancement models while significantly reducing computational complexity. These findings validate our bioacoustics-inspired design and establish BioSEN as an effective baseline for future research in the understudied field of animal vocalization enhancement.

\end{document}